\documentclass[12pt,a4paper]{article}
\usepackage[latin1]{inputenc}
\usepackage{amsmath}
\usepackage{amsfonts}
\usepackage{amssymb}
\usepackage{graphicx}
\usepackage{natbib}
\usepackage[left=1.00in, right=1.00in, top=1.00in, bottom=1.00in]{geometry}
\usepackage{setspace}
\usepackage{algorithm}
\usepackage{algpseudocode}
\usepackage{pifont}

\newcommand{\bbeta}{\mbox{\boldmath{$\beta$}}}

\def\bSigma{\mbox{\boldmath{$\Sigma$}}}

\newcommand{\balpha}{\mbox{\boldmath{${\alpha}$}}}

\def\smt{{\mbox{\tiny T}}}

\def\bY{\mathbf Y}
\def\bX{\mathbf X}
\def\bx{{\bf x}}
\def\by{{\bf y}}

\def\bS{\mathbf S}

\def\bu{\mathbf u}
\def\bv{{\bf v}}

\def\bA{\mathbf A}
\def\bB{\mathbf B}
\def\bU{\mathbf U}

\def\bI{\mathbf I}

\def\bK{\mathbf K}
\def\bR{\mathbf R}
\def\bP{\mathbf P}

\def\bV{{\bf V}}

\def\bD{\mathbf D}

\usepackage{color}
\usepackage{comment}
\usepackage{ulem}
\usepackage{soul}

\def\bSig\mathbf{\Sigma}

\usepackage[figuresright]{rotating}

\author{Sandra E. Safo, Shuzhao Li, Qi Long \\Department of Biostatistics and Bioinformatics\\
	Department of Medicine, Division of Pulmonary\\ Allergy and Critical Care Medicine\\
	Emory University, Atlanta, GA}
\title{Integrative analysis of transcriptomic and metabolomic data via sparse canonical correlation analysis with incorporation of biological information}

\date{}

\begin{document}




\maketitle
\label{firstpage}


\begin{abstract}
Integrative analyses of different high dimensional data types are  becoming increasingly popular.  Similarly, incorporating prior functional relationships among variables in data analysis has been a topic of increasing interest as it helps elucidate underlying mechanisms among complex diseases. In this paper, the goal is to assess association between transcriptomic and metabolomic data from a Predictive Health Institute (PHI) study including healthy adults at high risk of developing cardiovascular diseases. To this end, we develop statistical methods for identifying sparse structure in canonical correlation analysis (CCA) with incorporation of biological/structural information. Our proposed methods use prior network structural information among genes and among metabolites to guide selection of relevant genes and metabolites in sparse CCA, providing insight on the molecular underpinning of cardiovascular disease. Our simulations demonstrate that the structured sparse CCA methods outperform several existing sparse CCA methods in selecting relevant genes and metabolites when structural information is informative and are robust to mis-specified structural information. Our analysis of the PHI study reveals that a number of genes and metabolic pathways including some known to be associated with cardiovascular diseases are enriched in the subset of genes and metabolites selected by our proposed approach.
\end{abstract}

%
%

\textbf{Corresponding Author}: Sandra E. Safo; Email: ssafo@emory.edu


\section{Introduction \label{section intro}}
Recent advancement in high-throughput, biomedical technologies has enabled the measurement of multiple data types in the same studies, including genomics, epigenomics, transcriptomics and metabolomics.  Each of these data types provides a different snapshot
 of the underlying  biological system, and combining multiple data types has been shown to be very valuable in investigating complex diseases.  It has been demonstrated that individual components in these data are functionally structured in networks or pathways and incorporation of such structural information can improve analysis and lead to biologically more meaningful results \citep{LL:2008, PXS:2010, Chenetal:2013}. By the same token, it is desirable to jointly study the association between these data types with incorporation of available structural information for each data type, enabling us to uncover drivers that individually or in combination provide better insight about the biological mechanism. In this article, we develop new canonical correlation analysis (CCA) methods for studying the overall dependency structure  between transcripts and metabolites while incorporating structural information for each data type.

\subsection{The PHI Study}\label{sub:PD}
Our work is  motivated by data from the Emory  University and Georgia Tech PHI study. The PHI was established in 2005 with the goal of understanding and optimizing health focused on  maintaining health rather than treating disease.  
The PHI data are collected from a  longitudinal study of health measures in over $750$ healthy employees of Emory University and Georgia Tech. We use data for $52$ participants for whom gene expression and metabolomics data at baseline were available, 
and who were also at high risk of developing cardiovascular diseases defined by the Framingham risk scores \citep{DAgotinoetal:2008}.
The data consist of $32$ females and $20$ males with ages ranging from $19$ to $67$ years with a mean age of $47.35$ years. The gene expression data consist of $38,624$ probes and the metabolomic features  consist of $6,009$ features, where each metabolomic feature is defined by mass-to-charge ratio ($m/z$) and retention time and its relative concentration is captured by ion intensity \citep{RUPTJ:2014}. We pre-process the gene expression data using approaches from \citep{KKB:2003}. Specifically, we exclude genes with variance and entropy expression values that  are respectively less than the $90$th and $20$th percentile, resulting in $1,547$ genes. For the metabolomics data, we exclude features with more than $50\%$ zeros, and use  {mummichog} \citep{Li:2013} to annotate the $m/z$ features. This results in $252$ metabolites. 
 
Let $n=52$ be the common samples that have both transcriptomic and metabolomic data. We denote the trancriptomic and metabolomic data by $\bX=(\bx_{1},\cdots,\bx_{p})$ ($p=1,547$) and $\bY=(\by_{1},\cdots,\by_{q})$ ($q=252$), respectively, where $\bx, \by \in \Re^{n}$. Structural information for genes are represented by an undirected graph $\mathcal{G}_{X}=(C_{X},E_{X},W_{X})$, where $C_{X}$ is the set of nodes corresponding to the $p$ transcriptomic  features, $E_{X}=\{i\sim j\}$ is the set of edges indicating that features $i$ and $j$ are associated in a biologically meaningful way, and $W_{X}$ includes the weight of each node. Similarly, let $\mathcal{G}_{Y}=(C_{Y},E_{Y},W_{Y})$ be the structural information for metabolites. For node $i$ in $\bX$, denote by $d_{i}^{X}$ its degree i.e., the number of nodes that are directly connected to node $i$ and by $w_{i}^{X} = f(d_{i}^{X})$ its weight which can depend on $d_{i}^{X}$. Similarly, we define $d_{i}^{Y}$ and $w_{i}^{Y}$.  We use $w_{i}^{X}=d_{i}^{X}$ and $w_{i}^{Y}=d_{i}^{Y}$ in all our numerical studies. In our analysis of the PHI study, we obtain the gene network information from Kyoto Encyclopedia of Genes and Genomes (KEGG) \citep{KEGG:2016}, and the metabolomic network information from {mummichog} software \citep{Li:2013}. In the resulting gene network, there are $1,547$ genes with $479$ edges in total; the distribution of $d_{i}^{X}$ ranges from $1$ to $32$ with a mean of $3$. In the metabolomics network, there are $252$ metabolites and $190$ edges in total, with each edge representing  a connection between metabolites via a known metabolic reaction. The distribution of $d_{i}^{Y}$ for the metabolomics data ranges from $1$ to $13$ with a mean of $3$. 

Our goal is to assess association between genes and metabolites with incorporation of structural information for both data types, for which, to the best of our knowledge, little work has been done in statistical literature. It is especially challenging when the number of features ($p$ or $q$) greatly exceeds the sample size $n$ as the case in the motivating PHI  study, and in many biomedical omics studies. 

\subsection{Existing  Methods}
CCA was developed to find linear combinations of two sets of variables that have maximum correlation, which can help understand the overall dependency structure between these two sets of variables. However, it is well known that the classical CCA suffers from the singularity of sample covariance matrices when applied to high dimensional data; it also lacks biological interpretability especially when the number of variables is large. Extensions of CCA have been proposed to overcome these limitations. Some modifications  deal with the singularity of sample covariance matrices by applying a ridge-type regularization \citep{Vinod:1976, Safo:2014}, assuming sample covariance matrices are identity matrices \citep{WTH:2009,PTB:2009, CF:2012}, or have some structure such as sparsity, bandable or Toeplitz \citep{Chenetal:2013}. \cite{Chaoetal:2015} considered the sample covariances to be nuisance parameters and replaced their precision matrices with pseudo-inverses. The problem of biological interpretability has been tackled by assuming some coefficients are zero, implying that those variables do not contribute to the overall association between the two sets of variables \citep{WWZ:2008, PTB:2009, WTH:2009, CF:2012, Chenetal:2013, Chaoetal:2015}. \cite{CF:2012} used the CCA algorithm of \citet{PTB:2009} and compared several penalty functions such as lasso \citep{Tibshirani:1994}, elastic net \citep{ZH:2005}, SCAD \citep{FL:2001} and hard-thresholding. They concluded that elastic net and particularly SCAD achieve maximum correlation between the canonical correlation variables with more sparse canonical vectors. To achieve sparsity on the canonical vectors, \cite{Safo:2014}  imposed a $l_{\infty}$ constraint on a modified generalized eigenvalue problem arising from the CCA optimization problem while minimizing the $l_{1}$ norm of linear coefficients, which was motivated by Dantzig Selector \citep{Dantzig:2007}. 

Despite the success of the available sparse CCA methods, their main limitation is that they do not exploit structural information among variables that is available for biological data such as transcriptomic and metabolomic data. Using available structural information, one can gain better understanding and obtain biologically more meaningful results from CCA. This has been demonstrated in the setting of sparse regression analysis \citep{PXS:2010,LL:2008, KX:2009}. Recently, \cite{Chenetal:2013} incorporated  phylogenetic information from the bacterial taxa in CCA to study association between nutrient intake and human gut microbiome composition. We note that our work is different from the structured sparse CCA of \cite{ChenLiuCarbonell:2012}. In their work, they consider functional relationships among one data type and impose a group lasso penalty on the variables. Also, they do not utilize edge information among variables within pathways, which we do in the current paper. 

\subsection{Our Approach}
We propose two structured sparse CCA methods that impose smoothness penalties on canonical correlation vectors and also allow for incorporating structural information such as gene and metabolic pathways to guide selection of important metabolites, transcripts, and pathways.

Our work makes several contributions. First, the proposed methods enable us to conduct integrative analysis of transcriptomic and metabolic data that achieves variable selection and incorporates structural information for both data types, leading to biologically more meaningful results as evidenced in our data application. Second, we develop an efficient algorithm that can handle high dimensional problems. Third, our extensive simulations demonstrate that the performance of the proposed approach is similar to or better than several existing methods even when network structure is not informative for selection of important variables. In particular, our proposed methods offer several improvements over the recent work by \cite{Chenetal:2013}.  First, our CCA formulation comes from the generalized eigenvalue problem rather than the direct CCA optimization problem. This formulation is not only simple to understand, but it also allows us to use convex  objectives and constraints in the optimization problem that can be solved by most mathematical optimization softwares. Second, we use structural information from both sets of variables as opposed to only one set of variable, which is not a trivial extension. Third, their method and most sparse CCA methods assume that sample covariance matrices are identity matrices, but we relax this assumption as it can be overly restrictive in practice. In particular, our method allows the use of sparse covariance matrices \citep{FHTGL:2007,Yuan07modelselection:2007} from which the underlying structural network may be inferred.  

In section \ref{sec:methods}, we present the proposed structured sparse CCA after briefly reviewing sparse CCA. In Section \ref{sec:computations}, we present the algorithms for implementing the proposed sparse CCA. In Section \ref{sec:simul}, we conduct simulation studies to assess the performance of our methods in comparison with several existing methods. In Section \ref{sec:real}, we apply our approach to the PHI study. We conclude with some discussion remarks in Section \ref{sec:discussion}.

\section{Methods} \label{sec:methods}
Following the notation introduced in Section 1, suppose that we have two sets of random matrices, an $ n \times p$ matrix $\bX = (\bx_{1},\ldots,\bx_{p})$, and an $n \times q$ matrix $\bY =  (\by_{1},\ldots,\by_{q})$, both of which, without generality, are standardized to have column mean $0$ and variance $1$. CCA \citep{Hotelling:1936} finds projections $\balpha \in \Re^{p}$ and  $\bbeta \in \Re^{q}$ such that the correlation between linear combinations $\bX\balpha$ and $\bY\bbeta$ is maximized. Mathematically, CCA finds vectors $\balpha$ and $\bbeta$ that solve
\begin{equation}\label{eqn:cca}
\rho =\max_{\balpha, \bbeta} \mbox{corr}(\bX\balpha, \bY\bbeta) = \max_{\balpha,\bbeta}\frac{\balpha^{\smt}\bSigma_{xy}\bbeta}{\sqrt{\balpha^\smt \bSigma_{xx}\balpha} \sqrt{\bbeta^\smt \bSigma_{xx}\bbeta}}, \nonumber\\
\end{equation}
where $\bSigma_{xx}$, $\bSigma_{yy}$ and $\bSigma_{xy}$ are population covariance and  cross-covariance matrices.
The optimization problem is equivalent to solving
\begin{eqnarray} \label{eqn:ccaopt}
\max_{\balpha,\bbeta} \balpha^{\smt}\bSigma_{xy}\bbeta ~~~\mbox{subject to}~~ \balpha^{\smt}\bSigma_{xx}\balpha =1~~ \mbox{and~~}\bbeta^{\smt}\bSigma_{yy}\bbeta =1.
\end{eqnarray}
Using Lagrangian multipliers and some  algebra, one can show that problem (\ref{eqn:ccaopt}) results in a generalized eigenvalue (GEV) problem of the form

\begin{equation} \label{lagrang1}
\left[\begin{array}{cc}
	0 & \bSigma_{xy} \\
	\bSigma_{yx} & 0
\end{array}  \right]  \left[\begin{array}{c}
\balpha  \\
\bbeta
\end{array}  \right] = \rho \left[\begin{array}{cc}
\bSigma_{xx} & 0 \\
0 & \bSigma_{yy}
\end{array}  \right]\left[\begin{array}{c}
\balpha  \\
\bbeta
\end{array}  \right],
\end{equation}
which can be solved by applying the singular value decomposition (SVD) to the matrix
\begin{eqnarray}\label{eqn:svdK}
\bK = \bSigma_{xx}^{-1/2}\bSigma_{xy}\bSigma_{yy}^{-1/2} = (\bu_{1},\ldots,\bu_{k})\bD(\bv_{1},\ldots,\bv_{k})^{\smt}.
\end{eqnarray}
Here, $k$ is the rank of the matrix $\bK$, $\bu_{j}$ and $\bv_{j}$, ($j=1,\ldots,k$) are the $j$th left and right singular vectors of $\bK$, and $\bD$ is a diagonal matrix containing singular values $\lambda_{j}$ of $\bK$ ordered from the largest to the smallest. It follows that the optimal coefficients in the linear combinations of $\bX$ and $\bY$ are given by
\begin{eqnarray} \label{nonsparse}
\tilde{\balpha}_{j} = \bSigma^{-1/2}_{xx}\bu_{j}, ~\tilde{\bbeta}_{j} = \bSigma^{-1/2}_{yy}\bv_{j}.
\end{eqnarray}
The vectors $\tilde{\balpha}_{j}$ and $\tilde{\bbeta}_{j}$ are called the $j$th canonical correlation vectors for $\bX$ and $\bY$ respectively, and are nonsparse. The random variables $\bX\tilde{\balpha}_{j}$ and $\bY\tilde{\bbeta}_{j}$ are known as the $j$th canonical correlation variables, and $\tilde{\rho}_{j}= \lambda_{j}$ is the $j$th canonical correlation coefficient. Thus, the  optimal coefficients in the linear combination yielding maximum correlation between $\bX$ and $\bY$ is a rank one approximation of the matrix $\bK$. When data are available, one can replace the population matrices  $\bSigma_{xx}^{-1/2}\bSigma_{xy}\bSigma_{yy}^{-1/2}$ by the sample versions $\bS_{xx}^{-1/2}\bS_{xy}\bS_{yy}^{-1/2}$, which results in consistent estimators of $\balpha$ and $\bbeta$ for fixed dimensions $p, q$, and large sample size $n$.

When  $p$ is  greater than $n$, regularization is desirable in order to obtain interpretable solutions to the optimization problem (\ref{eqn:ccaopt}). Despite the success of the existing regularized CCA methods, their main drawbacks, when applied to the setting of our interest, include failure to take full advantage of prior biological knowledge, and reliance on the assumption that $\bS_{xx} = \bI$, $\bS_{yy} =\bI$ which can be overly restrictive. Given the network information defined in Section~\ref{sub:PD}, we investigate two structured sparse CCA for incorporating prior biological information.

\subsection{Grouped Sparse CCA}\label{sec:GSCCA}
The first approach is the Grouped sparse CCA, similar in spirit with \citet{PXS:2010}. Utilizing the graph structure in section \ref{sub:PD}, we propose the following structured sparse CCA criterion that solves the GEV problem (\ref{lagrang1}): for the $k$th  ($k=1,\ldots K$) canonical correlation vector we  solve iteratively until convergence the following optimization problem
\begin{small}
\begin{eqnarray}\label{eqn:groupssca}
\min_{\balpha} \left\{(1-\eta)\sum_{i \sim j}\left(\frac{|\alpha_{i}|^{\gamma}}{w_{i}^{X}} + \frac{|\alpha_{j}|^{\gamma}}{w_{j}^{X}}\right)^{1/\gamma} + \eta\sum_{d_{i}^{X}=0} |\alpha_{i}| \right\} &~\mbox{subject~to}~&
(A) \|\bS_{xy}\tilde{\bbeta}_{k} - \tilde{\rho}_{k}\tilde{\bS}_{xx}\balpha\|_{\infty} \leq \tau_{x_{1}}~ \nonumber\\
&&(B) \|\tilde{\bS}_{xx}^{-1}\bS_{xy}\tilde{\bbeta}_{k} - \tilde{\rho}_{k}\balpha\|_{\infty} \leq \tau_{x_{2}}~ \nonumber\\
\min_{\bbeta} \left\{(1-\eta)\sum_{i \sim j}\left(\frac{|\beta_{i}|^{\gamma}}{w_{i}^{Y}} + \frac{|\beta_{j}|^{\gamma}}{w_{j}^{Y}}\right)^{1/\gamma} + \eta\sum_{d_{i}^{Y}=0} |\beta_{i}| \right\} &~\mbox{subject~to}~&
(A)\|\bS_{yx}\tilde{\balpha}_{k} - \tilde{\rho}_{k}\tilde{\bS}_{yy}\bbeta\|_{\infty} \leq \tau_{y_{1}} \nonumber\\
&&(B) \|\tilde{\bS}_{yy}^{-1}\bS_{yx}\tilde{\balpha}_{k} - \tilde{\rho}_{k}\bbeta\|_{\infty} \leq \tau_{y_{2}}\nonumber\\
\end{eqnarray}
\end{small}
where for some random vector $\bx \in \Re^{p}$, $\|\bx\|_{\infty}$ is the $l_{\infty}$ norm and is defined as $\max_i |x_{i}|, i=1,\dots,p$, $\tau_{x_{1}} >0$ and  $\tau_{y_{1}} >0$ are tuning parameters, $\gamma >1$ and $0\leq \eta < 1$ are fixed, and $\tilde{\balpha}_{k}$ and $\tilde{\bbeta}_{k}$ are the $k$th nonsparse canonical vectors defined in  (\ref{nonsparse}).
As defined,  (A) and (B) represent two different sets of constraints and are discussed in detail in Section~\ref{subsec:computations}. The first term  in each objective function is the weighted grouped penalty \citep{PXS:2010}, which induces grouped variable selection. It encourages both $\alpha_{i}$ and $\alpha_{j}$ (similarly both $\beta_{i}$ and $\beta_{j}$ ) to be equal to zero or nonzero simultaneously, implying that two neighboring variables in a network are more likely to (or not to ) participate in the same biological process simultaneously. 
In addition, the weight $w_{i}^{X}$ encourages $|\alpha_{i}|/w_{i}^{X} =|\alpha_{j}|/w_{j}^{X}$  (similarly $|\beta_{i}|/w_{i}^{Y} =|\beta_{j}|/w_{j}^{Y}$) for two neighboring nodes $i$, $j$, allowing for connected features to have opposite effects. The second term in each objective function encourages variable selection of singletons that are not connected to any variable in the network. The tuning parameters  $\tau_{x_{1}}$ or  $\tau_{x_{2}}$ and  $\tau_{y_{1}}$ or $\tau_{y_{2}}$ control the number of coefficients that are exactly zero with larger values encouraging more sparsity. The selection of $\tau_{x}$ and  $\tau_{y}$ is usually data-driven, and is discussed later.

We can find $\hat{\balpha}_{k}$ and $\hat{\bbeta}_{k}$, $k \ge 2$ by solving (\ref{eqn:groupssca}) after projecting data onto the orthogonal complement of $[\hat{\balpha}_{1},\ldots,\hat{\balpha}_{k-1}]$ and $[\hat{\bbeta}_{1},\ldots,\hat{\bbeta}_{k-1}]$ respectively. In other words, we deflate data by obtaining $\bX_{new}= \bX{\bP^{\perp}_{k}}$, where $\bP^{\perp}_{k}$ is the projection matrix onto the othorgonal complement of $[\hat{\balpha}_{1},\ldots,\hat{\balpha}_{k-1}]$. We obtain $\bY_{new}$ similarly. 


In addition, in most of the existing sparse CCA methods,  $\bS_{xx}$ (and $\bS_{yy}$) is assumed to be an identity matrix, essentially assuming that $\bX$ (and $\bY$) is independent. We replace this assumption with the following variance-covariance matrices in our optimization problems
\begin{equation}\label{eqn:covdef}
\tilde{\bS}_{xx}=\bS_{xx} + \sqrt{\log{p}/n}\bI, ~~\tilde{\bS}_{yy} =\bS_{yy} + \sqrt{\log{q}/n}\bI
\end{equation}
similar in spirit with \cite{Vinod:1976}.
The optimization problems in (\ref{eqn:groupssca}) are convex and can be solved with an off-the-shelf convex optimization package such as the CVX package in Matlab. We provide remarks on merits of constraints (A) and (B) in Section~\ref{sec:computations}. Since the proposed method uses the nonsparse solution ($\tilde{\balpha}_{k}, \tilde{\bbeta}_{k}, \tilde{\rho}_{k})$ as the `initial' values, it is possible that the  effectiveness of the proposed method can be dependent on the quality of initial values. To alleviate the dependence we propose to iterate the procedure by updating the $(\tilde{\balpha}_k, \tilde{\bbeta}_k, \tilde{\rho}_k)$ with the found $(\hat{\balpha}_k, \hat{\bbeta}_k, \hat{\rho}_k)$ until convergence. Here $\hat{\rho}_k$ is the correlation coefficient between $\bX\hat{\balpha}_k$ and $\bY\hat{\bbeta}_k$.  Algorithm 1 below describes the procedure to obtain $\hat{\balpha}_k$ and $\hat{\bbeta}_k$, $k = 1, \ldots, K$.

\subsection{Fused Sparse CCA}
The second structured sparse CCA is the Fused sparse CCA, similar in spirit with \citet{TSRZK:2005}. Utilizing the graph structure $\mathcal{G}$ in section \ref{sub:PD}, we propose the following structured sparse CCA criterion that solves the GEV problem (\ref{lagrang1}): for the $k$th  ($k=1,\ldots K$) canonical correlation vector we  solve iteratively until convergence the following optimization problem
\begin{eqnarray} \label{fuseddz}
\min_{\balpha}\left\{(1-\eta)\sum_{i\sim j}\left|\frac{\alpha_{i}}{w_{i}^{X}}- \frac{\alpha_{j}}{w_{j}^{X}}\right| + \eta\sum_{d_{i}^{X}=0}|\alpha_{j}|\right\} &~\mbox{subject~to}~&
(A) \|\bS_{xy}\tilde{\bbeta}_{k} - \tilde{\rho}_{k}\tilde{\bS}_{xx}\balpha\|_{\infty} \leq \tau_{x_{1}}~ \nonumber\\
&&(B) \|\tilde{\bS}_{xx}^{-1}\bS_{xy}\tilde{\bbeta}_{k} - \tilde{\rho}_{k}\balpha\|_{\infty} \leq \tau_{x_{2}}~ \nonumber\\
\min_{\bbeta}\left\{(1-\eta)\sum_{i\sim j}\left|\frac{\beta_{i}}{w_{i}^{Y}}- \frac{\beta_{j}}{w_{j}^{Y}}\right| + \eta\sum_{d_{i}^{Y}=0}|\beta_{j}|\right\} &~\mbox{subject~to}~&
(A)\|\bS_{yx}\tilde{\balpha}_{k} - \tilde{\rho}_{k}\tilde{\bS}_{yy}\bbeta\|_{\infty} \leq \tau_{y_{1}} \nonumber\\
&&(B) \|\tilde{\bS}_{yy}^{-1}\bS_{yx}\tilde{\balpha}_{k} - \tilde{\rho}_{k}\bbeta\|_{\infty} \leq \tau_{y_{2}}\nonumber\\
\end{eqnarray}
where $\tau_{x_{1}} >0$ and  $\tau_{y_{1}} >0$ are tuning parameters, $0\leq \eta < 1$ is assumed fixed, and $\tilde{\balpha}_{k}$ and $\tilde{\bbeta}_{k}$ are the $k$th nonsparse canonical vectors defined in  (\ref{nonsparse}). (A) and (B) are the same two sets of constraints introduced in Section~\ref{sec:GSCCA}. This penalty is a combination of fused lasso penalty on variable pairs that are connected in the network and an $l_{1}$ penalty on singletons that are not connected to any other variable in the network. 
This penalty is  similar to the network constrained penalty of \citet{LL:2008}, but different in a number of ways. Their penalty
\begin{equation}
\eta_{1}\sum_{j}|\alpha_{j}| + \eta_{2}\sum_{i\sim j}\left(\frac{\alpha_{i}}{w_{i}}- \frac{\alpha_{j}}{w_{j}}\right)^2 \nonumber\\
\end{equation}
uses the $l_{2}$ norm and it has been shown that this does not produce sparse solutions, where sparsity refers to variables that are connected in a network. In other words, it does not encourage grouped selection of variables in the network \citep{PXS:2010}. In addition, the penalty $\eta_{2}\sum_{i\sim j}\left(\frac{\alpha_{i}}{w_{i}}- \frac{\alpha_{j}}{w_{j}}\right)^2$ produces a ``wigly" solution that is less attractive for interpretation \citep{TSRZK:2005}. On the other hand, the penalty $(1-\eta)\sum_{i\sim j}\left|\frac{\alpha_{i}}{w_{i}}- \frac{\alpha_{j}}{w_{j}}\right|$ gives a piecewise constant solution and can be interpreted as a simple weighted average of features that are connected in a network. Also, the additional tuning parameter $\eta_{2}$ introduces more computational costs when applied to CCA as done in \citet{Chenetal:2013}; it requires solving a graph-constrained regression problem with dimension $(n+p) \times p$, incurring a high computational cost for very large $p$, particularly if one incorporates structural information on $\bY$ as well.
Again, we  replace $\bS_{xx}$ and $\bS_{yy}$ by $\tilde{\bS}_{xx}$ and  $\tilde{\bS}_{yy}$ respectively.

\section{Computation and Algorithms} \label{sec:computations}
\subsection{Computations}\label{subsec:computations}
Of the two constraints in optimization problems~\eqref{eqn:groupssca} and \eqref{fuseddz}, constraint (B) is computationally motivated. Let $\hat{\balpha}_{F}$ and $\hat{\bbeta}_{F}$ be solution vectors from the structured sparse optimization with constraint (A) and let $\hat{\balpha}_{S}$, $\hat{\bbeta}_{S}$ be solution vectors from  constraint (B). It is straightforward to show that if $\tau_{x_{1}}=0$, $\tau_{y_{1}}=0$ and $\tau_{x_{2}}=0$, $\tau_{y_{2}}=0$, then  $\hat{\balpha}_{F}= \hat{\balpha}_{S}$ and  $\hat{\bbeta}_{F}= \hat{\bbeta}_{S}$, that is, the solution vectors are the same. However, for $\tau_{x_{1}}> 0, \tau_{y_{1}} >0, \tau_{x_{2}} >0$ and $\tau_{y_{2}} >0$, the optimization problems may yield the same objective functions but the solution vectors may not be the same, i.e., $\hat{\balpha}_{F} \ne \hat{\balpha}_{S}$ and  $\hat{\bbeta}_{F} \ne \hat{\bbeta}_{S}$.

When $p$ and $q$ are large, the optimization problems \eqref{eqn:groupssca} and \eqref{fuseddz} with constraint (A) are  expensive to compute using the CVX package since it requires inverting  $\bS_{xx}$, a $p \times p$ matrix, and $\bS_{yy}$, a $q \times q$ matrix, at each iteration. For constraint (B), a computationally efficient approach for very high dimensional problems is described as follows.
 Let
\begin{eqnarray*} \label{svdx}
\bX &=& \bU_{x}\bD_{x}\bV_{x}^{\smt}\nonumber\\
 &=& \bR_{x}\bV_{x}^{\smt}
\end{eqnarray*}
be the SVD of $\bX$, where $\bV_{x}$ is a $p \times n$ matrix of right singular vectors with orthonormal columns, $\bU_{x}$ is an $n \times n$ orthogonal matrix of left singular vectors and $\bD_{x}$ is a diagonal matrix of singular values. Hence $\bR_{x} = \bU_{x}\bD_{x}$ is also $n \times n$. Also let
 \begin{eqnarray*} \label{svdy}
\bY &=& \bU_{y}\bD_{y}\bV_{y}^{\smt}\nonumber\\
&=& \bR_{y}\bV_{y}^{\smt}
\end{eqnarray*}
be the SVD of $\bY$, where  $\bV_{y}$ is a $q \times n$ orthonormal matrix, $\bU_{y}$ is a $n \times n$ orthogonal matrix and  $\bD_{y}$ is a diagonal matrix of singular values. Then $\bR_{y} = \bU_{y}\bD_{y}$ is also $n \times n$. Plugging these into $\tilde{\bS}_{xx}^{-1}\bS_{xy}$, and after some careful linear algebra, we obtain
\begin{equation}
\tilde{\bS}_{xx}^{-1}\bS_{xy} = (\bS_{xx} + \sqrt{\log{p}/n}\bI)^{-1}\bS_{xy} = \bV_{x}(\bR_{x}^{\smt}\bR_x + \sqrt{\log{p}/n}\bI)^{-1}\bR_{x}^{\smt}\bR_{y}\bV_{y}^{\smt}, \nonumber
\end{equation}
which requires the inversion of an $n \times n$ matrix. Similarly,
\begin{equation}
\tilde{\bS}_{yy}^{-1}\bS_{yx}=(\bS_{yy} + \sqrt{\log{q}/n}\bI)^{-1}\bS_{yx} = \bV_{y}(\bR_{y}^{\smt}\bR_y + \sqrt{\log{q}/n}\bI)^{-1}\bR_{y}^{\smt}\bR_{x}\bV_{x}^{\smt}. \nonumber
\end{equation}
The same idea can be used in  \eqref{eqn:svdK} and \eqref{nonsparse} for the nonsparse estimates $\tilde{\balpha}_{k}$ and $\tilde{\bbeta}_{k}$ in both constraints (A) and (B) to reduce computational cost of obtaining SVD of  a $p \times q$ matrix,  which is expensive as $\min(p,q)$ increases.

\subsection{Algorithms}\label{subsec:algorithm}
We describe two algorithms for the proposed structured sparse CCA methods. The first algorithm obtains the $k$th canonical correlation vector for fixed tuning parameters $\tau_{x}$ and $\tau_{y}$. The second algorithm provides a data driven approach for selecting the optimal tuning parameters.

We first normalize the columns of $\bX$ and $\bY$ to have mean zero and unit variance. Let $\bu_{k}$ and $\bv_{k}$ be the $k$th left and right singular vectors of $\tilde{\bS}_{xx}^{-1/2}{\bS}_{xy}\tilde{\bS}_{yy}^{-1/2}$, and let $\lambda_{k}$ be the $k$th singular value. The approach discussed in Section~\ref{subsec:computations} can be used here for problems with large $p$ and/or $q$.
For  fixed positive tuning parameters $\tau_{x}$ and $\tau_{y}$, use Algorithm \ref{alg:ccavectors} for the $k$th sparse canonical correlation vectors, $\hat{\balpha}_{k}$ and $\hat{\bbeta}_{k}$.

The tuning parameters $\tau=(\tau_{x}, \tau_{y})$ control the model complexity and their optimal values need to be selected. We use $V$-fold cross validation (CV) to select $\tau$ at each iteration of Algorithm \ref{alg:ccavectors}. The optimal tuning parameter pair is  chosen by performing a grid search over the entire pre-specified set of parameter values. To further reduce computational costs, we use a cross search over the pre-specified set of parameters. For a fixed value in the $\tau_{y}$ set of values (we fix $\tau_{y}$ as the middle value of the set of values), we search over the entire space of $\tau_x$ values and select $\tau_{x_\text{opt}}$ that minimizes criterion (\ref{eqn:tunsel2}) given $\tau_{y}$. Using $\tau_{x_\text{opt}}$, we search the entire $\tau_{y}$ space and choose $\tau_{y_\text{opt}}$ that also minimizes criterion (\ref{eqn:tunsel2}). We choose $\tau_\text{opt} = (\tau_{x_\text{opt}}, \tau_{y_\text{opt}})$ at each iteration in Algorithm $1$ since the selected optimal pair from previous iterations may be too large and may result in a trivial solution at the subsequent iteration.

\begin{algorithm}
\caption{Optimization for obtaining the $k$th structured sparse CCA vector} \label{alg:ccavectors}
\begin{algorithmic}[1]
\For{ $k=1,\ldots,K$}  
\State Initialize  with nonsparse estimates: $\tilde{\balpha}_{k0}=\tilde{\bS}^{-1/2}_{xx}\bu_{k}$, $\tilde{\bbeta}_{k0}=\tilde{\bS}^{-1/2}_{yy}\bv_{k}$ with unity $l_{2}$ norm, and $\tilde{\rho}_{k0}= \lambda_{k}^{1/2}$. The approach discussed in Section~\ref{subsec:computations} can be used here for problems with large $p$ and/or $q$.
\For{t =1 until convergence or some maximum number of iterations}
\State Solve one of the following two optimization problems using  previous estimates
$\hat{\balpha}_{k(t-1)}$ and $\hat{\bbeta}_{k(t-1)}$, to obtain the $k$th estimates $\hat{\balpha}_{k(t)}$ and $\hat{\bbeta}_{k(t)}$:
\begin{enumerate}
\item[(3i)] The Grouped sparse optimization problem
\begin{scriptsize}
\begin{eqnarray*}
\min_{\balpha} \left\{(1-\eta)\sum_{i \sim j}\left(\frac{|\alpha_{i}|^{\gamma}}{w_{i}^{X}} + \frac{|\alpha_{j}|^{\gamma}}{w_{j}^{X}}\right)^{1/\gamma} + \eta\sum_{d_{i}^{X}=0} |\alpha_{i}| \right\} &~\mbox{subject~to}~&
(A)\|\bS_{xy}\hat{\bbeta}_{k(t-1)} - \hat{\rho}_{k(t-1)}\tilde{\bS}_{xx}\balpha\|_{\infty} \leq \tau_{x_{1}}~
\nonumber\\
&&(B) \|\tilde{\bS}_{xx}^{-1}\bS_{xy}\hat{\bbeta}_{k(t-1)} - \hat{\rho}_{k(t-1)}\balpha\|_{\infty} \leq \tau_{x_{2}}~ \nonumber\\
\min_{\bbeta} \left\{(1-\eta)\sum_{i \sim j}\left(\frac{|\beta_{i}|^{\gamma}}{w_{i}^{Y}} + \frac{|\beta_{j}|^{\gamma}}{w_{j}^{Y}}\right)^{1/\gamma} + \eta\sum_{d_{i}^{Y}=0} |\beta_{i}| \right\} &~\mbox{subject~to}~&
(A)\|\bS_{yx}\hat{\balpha}_{k(t-1)} - \hat{\rho}_{k(t-1)}\tilde{\bS}_{yy}\bbeta\|_{\infty} \leq \tau_{y_{1}} \nonumber\\
&&(B) \|\tilde{\bS}_{yy}^{-1}\bS_{yx}\hat{\balpha}_{k(t-1)} - \hat{\rho}_{k(t-1)}\bbeta\|_{\infty} \leq \tau_{y_{2}}~ \nonumber\\
\end{eqnarray*}\item[(3ii)] The Fused sparse optimization problem
\begin{eqnarray*}
\min_{\balpha}\left\{(1-\eta)\sum_{i\sim j}\left|\frac{\alpha_{i}}{w_{i}^{X}}- \frac{\alpha_{j}}{w_{j}^{X}}\right| + \eta\sum_{d_{i}^{X}=0}|\alpha_{j}|\right\} &~\mbox{subject~to}~&
(A)\|\bS_{xy}\hat{\bbeta}_{k(t-1)} - \hat{\rho}_{k(t-1)}\tilde{\bS}_{xx}\balpha\|_{\infty} \leq \tau_{x_{1}}~
\nonumber\\
&&(B)\|\tilde{\bS}_{xx}^{-1}\bS_{xy}\hat{\bbeta}_{k(t-1)} - \hat{\rho}_{k(t-1)}\balpha\|_{\infty} \leq \tau_{x_{2}}~
\nonumber\\
\min_{\bbeta}\left\{(1-\eta)\sum_{i\sim j}\left|\frac{\beta_{i}}{w_{i}^{Y}}- \frac{\beta_{j}}{w_{j}^{Y}}\right| + \eta\sum_{d_{i}^{Y}=0}|\alpha_{j}|\right\} &~\mbox{subject~to}~&
(A)\|\bS_{yx}\hat{\balpha}_{k(t-1)} - \hat{\rho}_{k(t-1)}\tilde{\bS}_{yy}\bbeta\|_{\infty} \leq \tau_{y_{1}} \nonumber\\
&&(B) \|\tilde{\bS}_{yy}^{-1}\bS_{yx}\hat{\balpha}_{k(t-1)} - \hat{\rho}_{k(t-1)}\bbeta\|_{\infty} \leq \tau_{y_{2}}~ \nonumber\\
\end{eqnarray*}
\end{scriptsize}
\end{enumerate}
\State Normalize $\hat{\balpha}_{k(t)}$ and $\hat{\bbeta}_{k(t)}$ to have unity $l_{2}$ norm and obtain the canonical correlation coefficient $\hat{\rho}_{k(t)}$.
\State Update $(\tilde{\balpha}_{k}, \tilde{\bbeta}_{k}, \tilde{\rho}_{k})$ with $(\hat{\balpha}_{k}, \hat{\bbeta}_{k}, \hat{\rho}_{k})$.
\EndFor
\State If $k \ge 2, k \le \min(n-1,p,q)$,  update $\bX$ and $\bY$ by projecting them to the orthogonal complement of 
 $[\hat{\balpha}_{1},\ldots,\hat{\balpha}_{k-1}]$ and $[\hat{\bbeta}_{1},\ldots,\hat{\bbeta}_{k-1}]$ respectively, and repeat steps $3$ to $7$.
 \EndFor
\end{algorithmic}
\end{algorithm}

\begin{algorithm}
\caption{V-fold CV for tuning parameter selection} \label{alg:ccatune}
\begin{algorithmic}[2]
\State Randomly group the rows of $\bX$ and $\bY$ into $V$ roughly equal-sized groups, denoted by $
\bX^{1},\ldots,\bX^{V}$, and $\bY^{1},\ldots,\bY^{V}$, respectively.

\For{ each $\tau_{x}$ and a fixed $\tau_{y}$ }
\begin{enumerate}
\item[(i)]  For $v=1,\ldots,V$, let $\bX^{-v}$ and $\bY^{-v}$ be the data matrix leaving out $\bX^{v}$ and $\bY^{v}$ respectively. Apply Algorithm \ref{alg:ccavectors} on $\bX^{-v}$ and $\bY^{-v}$ to derive the desired number of canonical correlation vectors $\hat{\balpha}^{-v}_{k}(\tau_{x}, \tau_{y})$,  and $\hat{\bbeta}^{-v}_{k}(\tau_{x}, \tau_{y}), k=1,\cdots,\min(n-1,p,q)$.
\item[(ii)]  Project $\bX^{v}$ and $\bY^{v}$ onto $\hat{\balpha}^{-v}_{k}(\tau_{x}, \tau_{y})$,  and $\hat{\bbeta}^{-v}_{k}(\tau_{x}, \tau_{y})$
to obtain the testing correlation coefficients, $\hat{\rho}_{k_{test}}^{v} (\tau_{x}, \tau_{y})= \text{corr}( \bX^{v}\hat{\balpha}^{-v}_{k}, \bY^{v}\hat{\bbeta}^{-v}_{k})$.
\item[(iii)]  Project $\bX^{-v}$ and $\bY^{-v}$ onto $\hat{\balpha}^{-v}_{k}(\tau_{x}, \tau_{y})$,  and $\hat{\bbeta}^{-v}_{k}(\tau_{x}, \tau_{y})$ to obtain the training correlation coefficients, $\hat{\rho}_{k_{train}}^{-v}(\tau_{x}, \tau_{y}) = \text{corr}( \bX^{-v}\hat{\balpha}^{-v}_{k}, \bY^{-v}\hat{\bbeta}^{-v}_{k})$.
\item[(iv)] Calculate the $V$-fold CV score as the  difference between the average training and testing absolute correlation coefficients.
\begin{eqnarray}\label{eqn:tunsel2}
CV(\tau_{x},\tau_{y})=\left|\frac{1}{V}\sum\limits_{v=1}^{V}\left|\hat{\rho}_{k_{train}}^{v} (\tau_{x}, \tau_{y})\right|- \frac{1}{V}\sum\limits_{v=1}^{V}\left|\hat{\rho}_{k_{test}}^{-v} (\tau_{x}, \tau_{y})\right|\right|.
\end{eqnarray}
\item[(v)] Select the optimal tuning parameter $\tau_{x}$ as $\tau_{x_{\text{opt}}} = \min CV(\tau_{x},\tau_{y})$
\end{enumerate}
\EndFor
\For{$\tau_{x_{\text{opt}}}$ and each $\tau_{y}$}
\begin{enumerate}
\item[(i)] Repeat steps $2(i)$ to $2(iv)$
\item[(ii)] Select the optimal tuning parameter $\tau_{y_{\text{opt}}}$ as $\tau_{y_{\text{opt}}} = \min \{CV(\tau_{x_{\text{opt}}},\tau_{y})\}$
 \end{enumerate}
\EndFor
 \State Apply $\tau_{\text{opt}} = ( \tau_{x_{\text{opt}}}, \tau_{x_{\text{opt}}} )$ on the whole training data
 $\bX$, $\bY$ to obtain the optimal canonical vectors $\hat{\balpha}_{k}$, $\hat{\bbeta}_{k}$, and coefficients $\hat{\rho}_{k}$ at each iteration until convergence.
\end{algorithmic}
\end{algorithm}

\section{Simulations}\label{sec:simul}
We conduct simulations to assess the performance of the proposed methods in comparison with several existing sparse CCA methods.

\subsection{Simulation Set-up}
Two hundred Monte Carlo (MC) datasets are generated as follows. The first data type $\bX$ have $p$ variables and the second data type $\bY$ have $q$ variables, all drawn on the same samples with size $n=80$. $(\bX, \bY)$ are simulated from MVN$(\mathbf{0}, \bSigma$) with mean $\bf{0}$ and covariance $\bSigma$ partitioned  as
\begin{eqnarray}
\bSigma =\left(
\begin{array}{cc}
	\bSigma_{xx} & \bSigma_{xy} \nonumber\\
	\bSigma_{yx} & \bSigma_{yy} \nonumber\
\end{array} \right),
\end{eqnarray}
where $\bSigma_{xy}$ is the covariance between $\bX$ and $\bY$, and $\bSigma_{xx}$, $\bSigma_{yy}$ are respectively the covariance of $\bX$ and $\bY$  that describe the network structure in each data type.  Without loss of generality, we let the first $36$ variables form the networks in $\bX$ and $\bY$, where within each data type there are $6$ main variables, each connected to $5$ variables. 
The resulting network has $36$ variables and edges with a maximum degree of $5$, and $p-36$ and $q-36$ singletons in $\bX$ and $\bY$ respectively. Using the notation in Section \ref{sub:PD}, the graph structure is given by  $\mathcal{G}_{X}=\mathcal{G}_{Y}=\{C, E,W\}$, where $C=\{i,j \in p, q\}$, $E =\{ i \sim j | i, j =1,\cdots,36\}$, and $W=\{w_{i}|w_{i}=\text{degree of node}~  i, i=1,\cdots,36\}$. 
The network structure in each data type is captured by the covariance matrices
\begin{eqnarray}
\bSigma_{xx} = \left(
\begin{array}{cc}
\bar{\bSigma}_{36 \times 36} & \bf{0} \nonumber\\
\bf{0} & \bI_{p-36} \nonumber\
\end{array} \right),~~
\bSigma_{yy} = \left(
\begin{array}{cc}
\bar{\bSigma}_{36 \times 36} & \bf{0} \nonumber\\
\bf{0} & \bI_{q-36} \nonumber\
\end{array} \right),
\end{eqnarray}
where $\bar{\bSigma}$ is block diagonal with $6$ blocks of size $6$, between-block correlation $0$ and within each block there is a $5 \times 5$ compound symmetric submatrix with correlation $0.49$ describing the correlation structure of the connected variables. The correlation between a main and a connecting variable is $0.7$. The covariance between $\bX$ and $\bY$ is $\bSigma_{xy} = \rho\bSigma_{xx}\balpha\bbeta^{\smt}\bSigma_{yy}$, and $\balpha$ and $\bbeta$ are the true canonical correlation vectors and $\rho$ is the canonical correlation coefficient.


We consider four simulation  scenarios.
\noindent
\begin{enumerate}

\item Scenario one: \textit{All networks in $\bX$ are correlated with all networks in $\bY$}

In the first scenario, all $6$ networks in $\bX$ and $\bY$ are  associated and contribute to the correlation between the sets of variables, while the remaining singletons do not contribute to the correlation and thus have zero coefficients.. We generate the true canonical correlation vectors  $\balpha$ and $\bbeta$ as follows
\begin{scriptsize}
\begin{eqnarray*}
\left(-20, \frac{-20}{\sqrt{5}},\ldots,\frac{-20}{\sqrt{5}}, 20,\frac{20}{\sqrt{5}},\ldots,\frac{20}{\sqrt{5}}, -17,\frac{-17}{\sqrt{5}},\ldots,\frac{-17}{\sqrt{5}},17,\frac{17}{\sqrt{5}},\ldots,\frac{17}{\sqrt{5}},-10,\frac{-10}{\sqrt{5}},\ldots,\frac{-10}{\sqrt{5}}, 10,\frac{10}{\sqrt{5}},\ldots,\frac{10}{\sqrt{5}},0,\ldots, 0\right)  \nonumber\\
\end{eqnarray*}
\end{scriptsize}
and normalize such that $\balpha^{\smt}\bSigma_{xx}\balpha=1$ and $\bbeta^{\smt}\bSigma_{yy}\bbeta=1$. The canonical correlation coefficient $\rho$ is taken as $0.9$.

\item Scenario two: \textit{Two networks in $\bX$ and $\bY$ are correlated}

In the second scenario, only the first $2$ networks  in $\bX$ and $\bY$ contribute to the correlation structure between the sets of variables. The remaining networks and singletons do not contribute to the correlation between the two data types, even though within each data type, each network exhibit strong association between variables.  The true canonical correlation vectors $\balpha$ and $\bbeta$ are generated as
\begin{small}
	\begin{eqnarray*}
		\left(-20, \frac{-20}{\sqrt{5}},\ldots,\frac{-20}{\sqrt{5}}, 20,\frac{20}{\sqrt{5}},\ldots,\frac{20}{\sqrt{5}},0,\ldots, 0\right)  \nonumber\
	\end{eqnarray*}
\end{small}
and we normalize each to have $\balpha^{\smt}\bSigma_{xx}\balpha=1$ and $\bbeta^{\smt}\bSigma_{yy}\bbeta=1$. The canonical correlation $\rho$ is again taken as $0.9$.
\item Scenario three: \textit{Two orthogonal CCA vectors in $\bX$ and $\bY$}

In the third scenario, there are two orthogonal canonical correlation vectors $\bA=(\balpha_{1}, \balpha_{2})$ and $\bB=(\bbeta_{1}, \bbeta_{2})$ in $\bX$ and $\bY$ respectively that induce the correlation between $\bX$ and $\bY$. Specifically, there are four networks in $\balpha_{1}$, which are the first $24$ variables with nonzero loadings, and these are associated with the first $18$ variables (or $3$ networks) in $\bbeta_{1}$. The next $12$ variables, forming the remaining two networks are found in $\balpha_{2}$, and these are correlated with the next $3$ networks in $\bbeta_{2}$. Then, the covariance matrix between $\bX$ and $\bY$ is $\bSigma_{xy} = \bSigma_{xx}\bA\bD\bB^{\smt}\bSigma_{yy}$, where $\bD=\text{diag}(0.9, 0.6)$ is a diagonal matrix with diagonal values being the first and second canonical correlation coefficients. We normalize the vectors
\begin{small}
\begin{eqnarray*}
\balpha_{1}& =& \left(-20, \frac{-20}{\sqrt{5}},\ldots,\frac{-20}{\sqrt{5}}, 20,\frac{20}{\sqrt{5}},\ldots,\frac{20}{\sqrt{5}}, -17,\frac{-17}{\sqrt{5}},\ldots,\frac{-17}{\sqrt{5}},17,\frac{17}{\sqrt{5}},\ldots,\frac{17}{\sqrt{5}},0,\ldots, 0\right)  \nonumber\\
\balpha_{2}&=& \left(0,\ldots,0,17,\frac{17}{\sqrt{5}},\ldots,\frac{17}{\sqrt{5}},-17,\frac{-17}{\sqrt{5}},\ldots,\frac{-17}{\sqrt{5}},0,\ldots, 0\right) \nonumber\
\end{eqnarray*}
\end{small}
and
\begin{small}
\begin{eqnarray*}
\bbeta_{1} &=& \left(-20, \frac{-20}{\sqrt{5}},\ldots,\frac{-20}{\sqrt{5}}, 20,\frac{20}{\sqrt{5}},\ldots,\frac{20}{\sqrt{5}}, -17,\frac{-17}{\sqrt{5}},\ldots,\frac{-17}{\sqrt{5}},0,\ldots, 0\right)  \nonumber\\
\bbeta_{2} &=& \left(0,\ldots,0,17,\frac{17}{\sqrt{5}},\ldots,\frac{17}{\sqrt{5}},-10,\frac{-10}{\sqrt{5}},\ldots,\frac{-10}{\sqrt{5}},10,\frac{10}{\sqrt{5}},\ldots,\frac{10}{\sqrt{5}},0,\ldots, 0\right) \nonumber\
\end{eqnarray*}
\end{small}
to have $\balpha_{i}^{\smt}\bSigma_{xx}\balpha_{i}=1$, $\bbeta_{i}^{\smt}\bSigma_{yy}\bbeta_{i}=1, i=1,2$, $\balpha_{1}^{\smt}\bSigma_{xx}\balpha_{2}=0$, and $\bbeta_{1}^{\smt}\bSigma_{yy}\bbeta_{2}=0$.

\item Scenario four: \textit{Randomly selected features in $\bX$ and $\bY$ are correlated}

In the fourth scenario,  there are two networks and $12$ variables in $\bX$ and $\bY$ that are correlated. These two networks are the same as those in scenario two, but are randomly dispersed and are not necessarily the first $12$ variables in $\bX$ and $\bY$. We normalize the vectors to have
$\balpha^{\smt}\bSigma_{xx}\balpha=1$ and $\bbeta^{\smt}\bSigma_{yy}\bbeta=1$. The canonical correlation $\rho$ is taken as $0.9$. This setting assesses performance in cases where the structural information  is mis-specified or uninformative and sheds light on robustness of the proposed methods.

%
\end{enumerate}

In the analysis of each MC dataset, we fix $\eta =.5$ and set $\gamma = 2$ in the $L_{\gamma}$-norm penalty of the Grouped structured sparse CCA method. We consider the dimensions $(p,q)=(500,500)$ for all scenarios. We use  $5$-fold cross validation to select the optimal tuning parameters from criterion $\ref{eqn:tunsel2}$, and then obtain $\hat{\balpha}$ and $\hat{\bbeta}$ using the entire training set. 

We evaluate the proposed methods based on their ability to select relevant features while maximizing correlation between $\bX$ and $\bY$. The results are summarized in terms of sensitivity, specificity, and Matthew's correlation coefficient (MCC) which are defined as follows:
\begin{eqnarray}
\mbox{Sensitivity}&=&\frac{TP}{TP + FN}, ~ \mbox{Specificity}=\frac{TN}{TN + FP}\nonumber\\
\mbox{MCC} &=& \frac{TP \cdot TN - FP \cdot FN}{\sqrt {(TP + FN)(TN + FP)(TP + FP)(TN + FN)}}, \nonumber
\end{eqnarray}
where TP, FP, TN, and FN are true positives, false positives, true negatives, and false negatives, respectively. Of note,  MCC lies in the interval $[-1,1]$, with a value of $1$ corresponding to selection of all signal variables and no noise variables, a perfect selection.
A value of $-1$ implies that $FP=1$, $FN=1$ and  $TP=0$, $TN=0$, and a value of $0$ implies $TP=TN=FP=FN=0.5$.

\subsection{Simulation results}
We denote the proposed methods, Grouped and Fused structured sparse CCA as  Grouped$_{A}$, Grouped$_{B}$, and Fused$_{A}$, Fused$_{B}$ with subscripts $A$ and $B$ respectively indicating constraints $A$ and $B$ in (\ref{eqn:groupssca}) and (\ref{fuseddz}). We compare with the following sparse methods: sparse CCA (SCCA) \citep{PTB:2009}, penalized matrix decomposition CCA (PMD) \citep{WTH:2009}, sparse CCA with SCAD penalty (SCAD) \citep{CF:2012} and sparse estimation via linear programming for CCA (SELP)\citep{Safo:2014}.

\begin{figure}[htbp]
\begin{center}
	\begin{tabular}{cc}		
		\includegraphics[height = 2.0in]{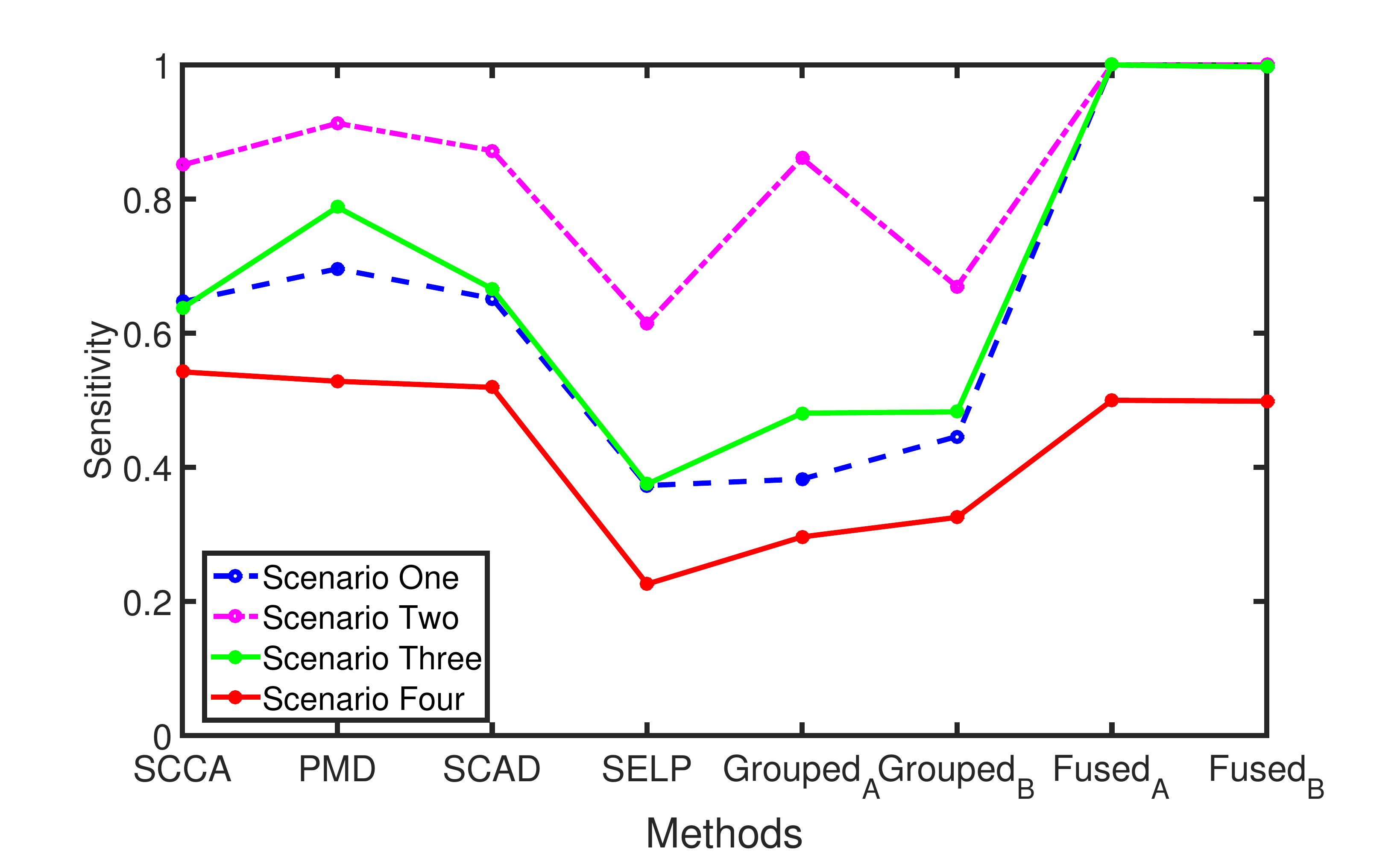}&\includegraphics[height = 2.0in]{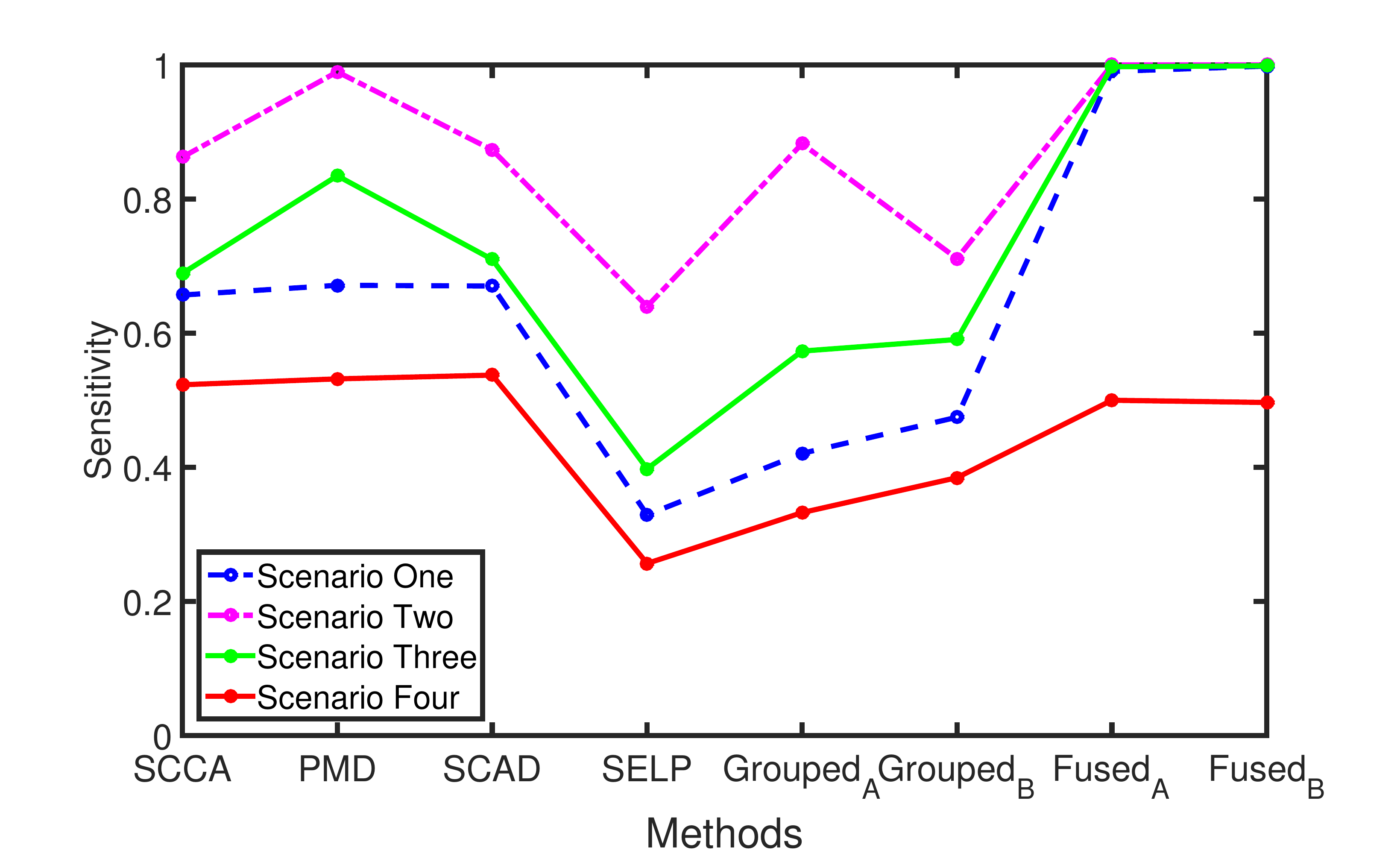}\\
~~~~~~~~~~~Sensitivity - $\bX$ &~~~~~~~~~~~	Sensitivity -$\bY$  \\
\includegraphics[height = 2.0in]{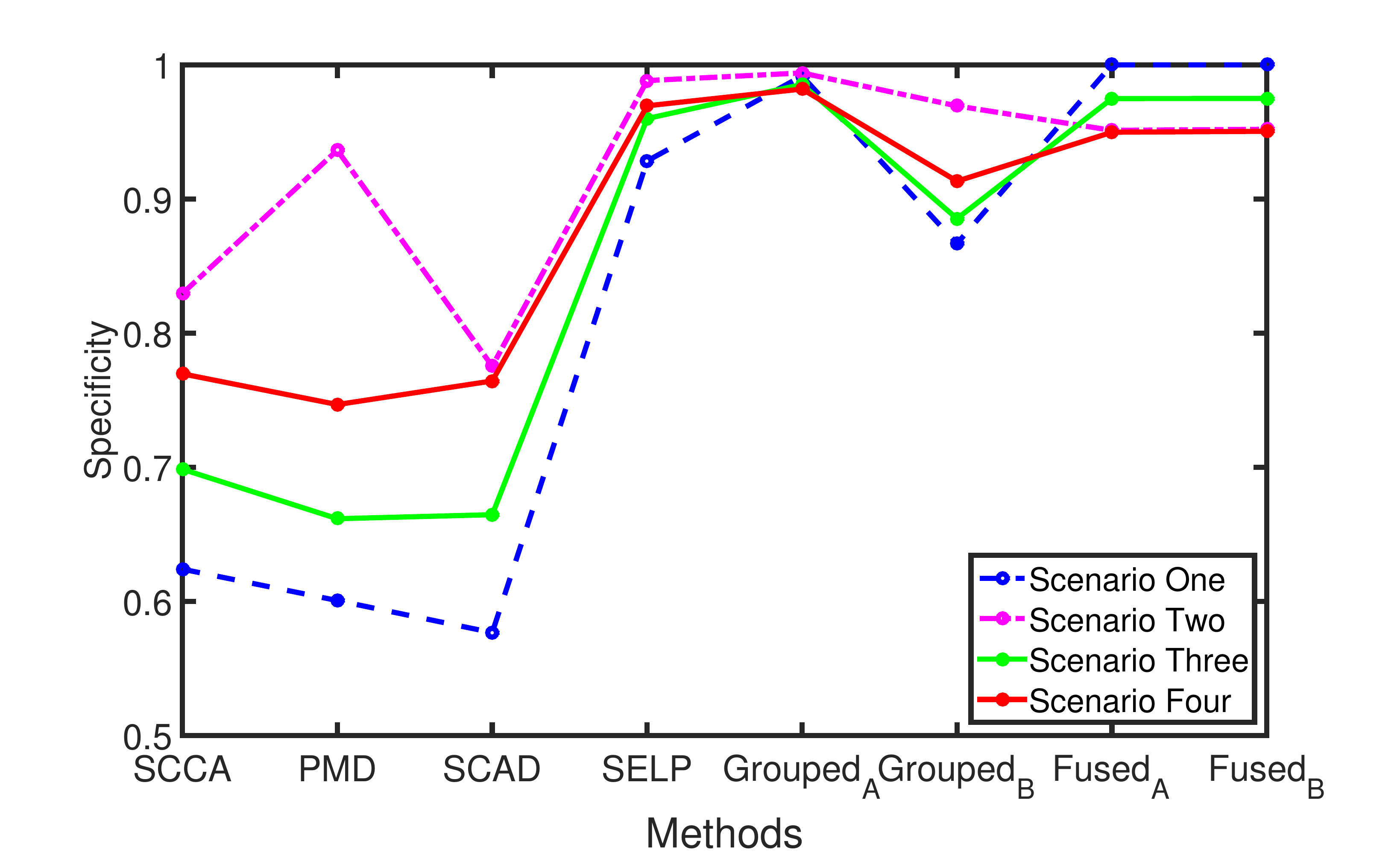}&\includegraphics[height = 2.0in]{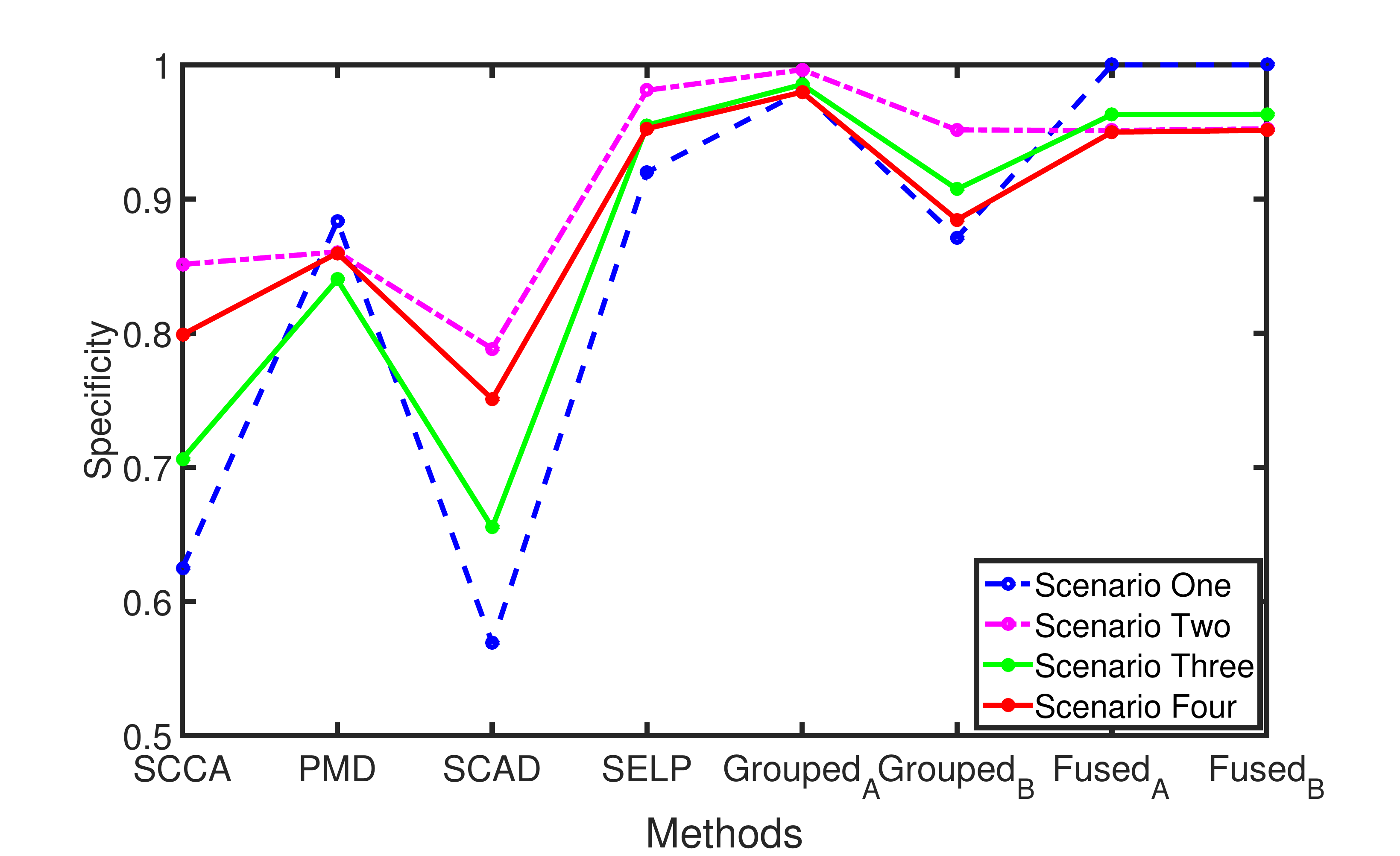}\\
~~~~~~~~~~~Specificity - $\bX$&~~~~~~~~~~~	Specificity -$\bY$  \\
\includegraphics[height = 2.0in]{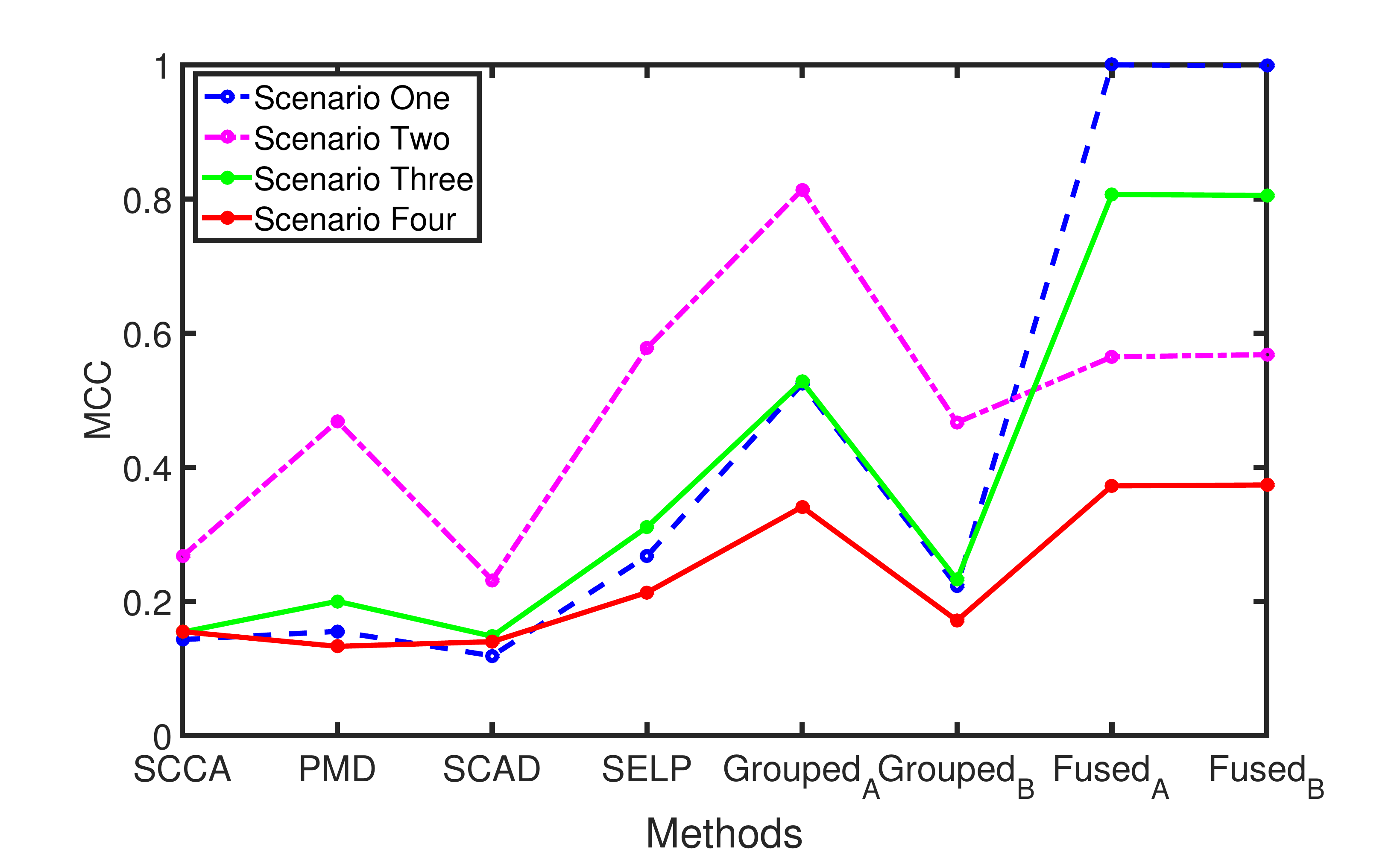}	&\includegraphics[height = 2.0in]{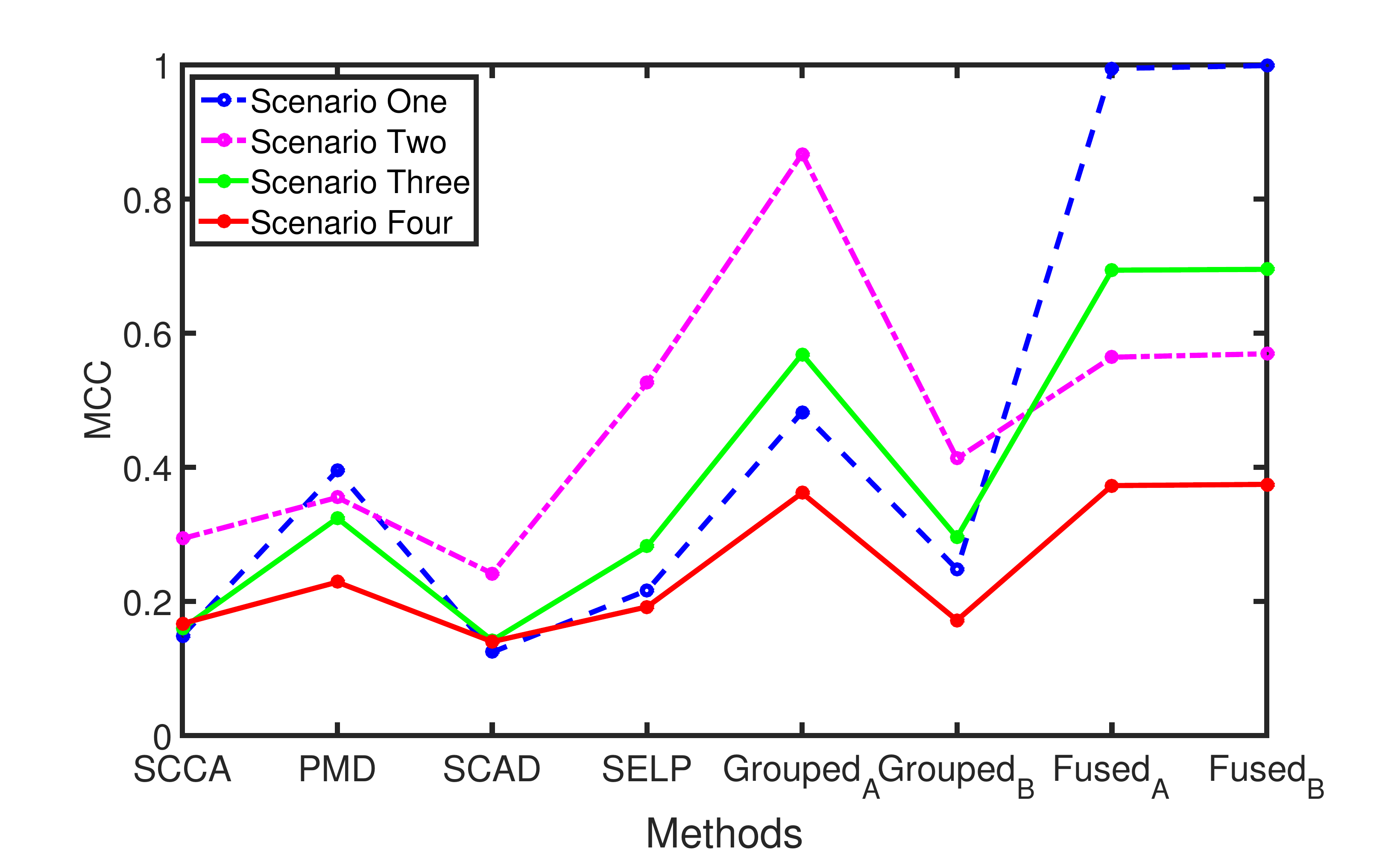}\\
~~~~~~~~~~~MCC- $\bX$ &~~~~~~~~~~~	MCC -$\bY$\\
\end{tabular}
\vspace{0.8cm}
\caption{\textbf{Comparison of structured sparse CCA with existing sparse CCA methods under scenarios one to four}. MCC, Matthew's correlation coefficient.
\label{fig:simulresults2}}
\end{center}
\end{figure}


Figure \ref{fig:simulresults2}  shows the sensitivity, specificity and MCC for the methods. We observe a competitive performance of the proposed methods, in particular Fused$_{A}$ and Fused$_{B}$, in selecting the true signals in all but scenario four. Fused$_{A}$ and Fused$_{B}$ perform well in scenarios one, three and four while Grouped$_{A}$ performs better in scenario four.  The other sparse methods especially SCCA and SCAD tend to select a large number of noise variables, evident by the low specificity and MCC proportions in Figure \ref{fig:simulresults2}.

For the proposed methods, it is noticeable from the sensitivity and MCC proportions that Grouped$_{A}$ and Grouped$_{B}$ have a suboptimal performance in scenarios one, two and three, yet these are better than the sparse methods. In scenario two, Fused$_{A}$ and Fused$_{B}$ select more FP  than Grouped$_{A}$ and Grouped$_{B}$ as evidenced by the low specificity, yet they are comparable to the other sparse methods. Recall that in scenario two, only $2$ networks in $\bX$ and $\bY$ contribute to the overall correlation between $\bX$ and $\bY$. However, within each network, there is high correlation, causing the Fused methods to read these as signals and therefore select them, though they do not contribute to the association between $\bX$ and $\bY$. In scenario four, the performance of all the methods deteriorates from scenarios one to three, yet the proposed methods still outperform the other sparse methods, suggesting that the proposed methods are robust to uninformative network information.

When we compare constraints A and B for Grouped and Fused methods, we notice similar performances in terms of variable selection and MCC for both Fused$_{A}$ and Fused$_{B}$, but the latter is  computationally more efficient and can be used for very high dimensional problems. For the Grouped method, Grouped$_{A}$ has  high specificity and high MCC values (Figure \ref{fig:simulresults2}), but Grouped$_{B}$ has better sensitivity. In general,   Grouped$_{A}$  outperforms Grouped$_{B}$ at higher computational cost.
Comparing the Fused and Grouped methods, we notice that in general, the performance of the Fused method is better than the Grouped method. However, the Grouped method (specifically Grouped$_{A}$) tends to achieve better performance in terms of specificity.

The results in Figure \ref{fig:simulresults2} demonstrate that the structured sparse CCA methods exhibit superior performance over the other sparse methods that are considered, evidenced by their high sensitivity, high specificity and high MCC proportions. The performance of the other sparse methods is worse in scenarios one and three than in scenario two. This shows that if the features in each set of variables are interconnected in the form of networks, and if most of these networks contribute to the association between $\bX$ and $\bY$, the existing sparse methods encounter difficulty in selecting the important networks. On the other hand, the proposed structured sparse methods can exploit the prior biological knowledge to increase sensitivity, specificity, and MCC.

\section{Analysis of the PHI Study Data} \label{sec:real}
We apply the proposed methods to integrative analysis of the transcriptomic $\bX$ and metabolomic $\bY$ data in the  PHI study. We $\log10$ transform the metabolomics data and normalize both the transcriptomic  and metabolomics data  to have mean $0$ and variance $1$ for each transcriptomic or metabolomic feature. Our goal  is to identify a subset of transcriptomic and metabolomic features that capture the overall association between transcripts and metabolites.


\indent
We apply the proposed methods and some existing sparse CCA methods to this PHI study. We use  $5$-fold cross validation to select optimal tuning parameters in our proposed methods, and then apply the selected tuning parameters to the whole data to estimate the maximal canonical correlation coefficient and vectors. Table \ref{tab:realresults3} shows the number of genes and metabolites from the first canonical correlation vectors. Table \ref{tab:realresults3b} shows the number of genes and metabolites that are common among the methods. From Table \ref{tab:realresults3}, we observe that the proposed methods, especially Grouped$_{B}$ and Fused$_{B}$ have high estimated canonical correlation coefficients compared to SELP even though all  select similar number of genes and metabolites. Of the proposed methods,  Grouped$_{A}$ is more sparse, which is consistent with the simulation results as observed by the low sensitivity and high specificity (Figure \ref{fig:simulresults2}) when compared with Fused$_{A}$, Grouped$_{A}$, and Grouped$_{B}$. In addition, the genes and metabolites identified by Fused$_{A}$ are subsets of those identified by Fused$_{B}$. It is noticeable in Table \ref{tab:realresults3b}  that there is considerable overlap of the genes and metabolites identified by the proposed methods and the existing methods considered. \\ 

\begin{table}[htbp]
	\caption{Number of genes and metabolites selected in the first  and second canonical correlation vectors in the PHI study. NA indicates that the underlying method only produce first canonical correlation vectors and coefficients. \label{tab:realresults3}}	
	\begin{center}
		\begin{small}
			\begin{tabular}{lccccc}
				\hline
				~		& Genes Selected &	 Metabolites selected&	Correlation Coefficient\\
				~		& $\hat{\balpha}_{1}$ &	 $\hat{\bbeta}_{1}$ &	$\hat{\rho}_{1}$ \\						
				\hline
				SCCA        	&86&154&	0.7248\\
				PMD            	&654&36&	0.8745\\
				SCAD       &31&252&0.7036\\
				SELP              &508&152&	0.8982\\		
				Grouped$_{A}$	        &9&4&	0.8168\\
				Grouped$_{B}$      &535&137&	0.9871\\	
				Fused$_{A}$       	&297&146& 	0.8658\\
				Fused$_{B}$      	&536&168&0.9814\\
				\hline
				\hline
			\end{tabular}
		\end{small}
	\end{center}
\end{table}

\begin{table}[htbp]
	\caption{Overlapping genes and metabolites selected in the first canonical correlation vectors in the PHI study. $(\cdot,\cdot)$ represents number of genes and metabolites common for each pair of method compared.\label{tab:realresults3b}}	
	\begin{center}
		\begin{scriptsize}
			\begin{tabular}{lcccccccc}
				\hline
				~		& SCCA&	PMD&SCAD&SELP&Grouped$_{A}$&Grouped$_{B}$&Fused$_{A}$&Fused$_{B}$\\
				\hline
				SCCA     &(86,154) &~&~&~& ~&~&~&~\\
				PMD       &(14,20)&(654,36)&~&~&~&~&~&~\\    	
				SCAD     &(31,154)&(10,36)&(31,252)&~&~&~&~&~\\
				SELP     &(16,92)&(503,33)&(11,152)&(508,152)&~&~&~&~\\     		
				Grouped$_{A}$	 &(0,2)&(9,3)&(0,4)&(9,3)&(9,4)&~&~&~\\    
				Grouped$_{B}$  &(17,82)&(427,30)&(9,137)&(383,101)&(9,3)&(535,137)&~&~\\  
				Fused$_{A}$ &(13,89)&(124,24)&(4,146)&(91,91)&(2,4)&(92,77)&(297,146)&~\\      	
				Fused$_{B}$ &(21,104)&(342,31)&(8,168)&(297,108)&(9,4)&  (323,98)&  (297,146)&(536,168)\\	
				\hline
			\end{tabular}
		\end{scriptsize}
	\end{center}
\end{table}

\indent We also investigate the biological relationships between the selected genes and metabolites using ToppGene Suite \citep{TopGene:2009} and MetaboAnalyst $3.0$ \citep{Xia:20042015} respectively. These genes and metabolites are taken as input in ToppGene and MetaboAnalyst $3.0$ online tools to identify pathways that are significantly enriched. The pathways that are significantly enriched in the genes selected by Fused$_{B}$ include mitochondrial ATP synthesis coupled proton transport and Oxidative phosphorylation. For the metabolites, the pathways identified in Fused$_{B}$ include purine and histidine metabolism.  These pathways play essential roles in some important biological processes including orderly cell division, cell proliferation, differentiation and migration, and survival. For instance, cardiovascular research suggests that oxidative phosphorylation is implicated in mitochondrial dysfunction, a major factor in heart failure \citep{Torstenetal:2013,Marinaetal:2008}. In addition, several epidemiological research suggest that uric acid, which is the final end product of purine metabolism \citep{Maiuolo:2015}, is an important and independent risk factor for cardiovascular diseases \citep{Fang:2000,Aiyer:2004} 

In conclusion, our analyses demonstrate that the proposed structured sparse CCA methods lead to biologically meaningful results that may shed light on the etiology of cardiovascular diseases.


\section{Discussion}\label{sec:discussion}
In this paper, we propose a new approach for integrative analysis of transcriptomic and metabolomic data. The two proposed methods, Grouped and Fused sparse CCA, allow us to not only assess association between two data types using  a subset of relevant genes and metabolites, but also take into account structural information from each data type. Simulation studies demonstrate that our methods achieve better performance than several other sparse CCA methods when prior network information is informative, and they are robust to mis-specified and uninformative network information. Applying the proposed approach to the PHI study, we show that a number of genes and metabolic pathways including some known to be associated with cardiovascular diseases are enriched in the subset of selected genes and metabolites that may shed light on the etiology of cardiovascular diseases.

Of the two methods proposed, our numerical studies show that the Fused sparse CCA performs better than the Grouped sparse CCA in terms of MCC and sensitivity, while the Grouped sparse CCA outperforms the Fused sparse CCA in terms of specificity. Our recommendation is  to use the Fused sparse CCA (particularly, Fused$_{B}$) for $p \gg n$ problems, and the Grouped sparse CCA (particularly, Grouped$_{A}$) for small to moderate dimensional problems. The proposed methods are implemented in MATLAB and are available upon request.
In the case where the graph information is not available, one can estimate network structures from observed data using existing approaches for sparse estimation of precision matrices \citep{FHTGL:2007, CLL:2011}.

While our current work has focused on continuous data,  it is of interest to develop similar methods for discrete data such as SNP data. When data are not continuous, CCA cannot be directly applied. To tackle this difficulty, one approach is to assume that there is a latent continuous variable for each discrete variable and use these latent variables to model the discrete variables where correlation among the latent variables is assessed using CCA.  It is also of interest to extend our methods to conduct integrative analysis of more than two data types and assess nonlinear associations between multiple omics data types.






%

\bibliographystyle{apalike} 
\bibliography{ssccabib}

\label{lastpage}

\end{document}